\documentclass[twocolumn,prb,showpacs,preprintnumbers,amsmath,amssymb]{revtex4} 
\usepackage{graphicx}
\usepackage{dcolumn}
\input epsf

\begin{document}

\title{Single-channel transmission in gold one-atom contacts and chains}
\author{G. Rubio-Bollinger$^1$, C. de las Heras$^1$, E. Bascones$^{2,3,*}$, N. Agra\" it$^1$, F. Guinea$^2$, S. Vieira$^1$ }

\affiliation{ $^1$ Instituto Universitario de Ciencia de
Materiales "Nicol\' as Cabrera",
 Universidad Aut\'onoma de Madrid, E-28049 Madrid, Spain.\\
$^2$Instituto de Ciencia de Materiales de Madrid, CSIC,
Cantoblanco, E-28049
Madrid, Spain \\
$^3$Departament of Physics, University of Texas at Austin, Austin,
TX 78712\\}


\begin{abstract}
We induce superconductivity by proximity effect in thin layers of
gold and study the number of conduction channels which contribute
to the current in one-atom contacts and atomic wires. The atomic
contacts and wires are fabricated with a Scanning Tunneling
Microscope. The set of transmission probabilities of the
conduction channels is obtained from the analysis of the $I(V)$
characteristic curve which is highly non-linear due to multiple
Andreev reflections. In agreement with theoretical calculations we
find that there is only one channel which is almost completely
open.

\end{abstract}

\maketitle

Much effort has been devoted in the last decade to the
understanding of electron transport processes and mechanical
properties of atomic-sized point contacts and chains between
metallic electrodes \cite{review}. Coherent electron transport in
these nanostructures can be understood in the frame of the
scattering formalism. The conductance $G$ of these nanocontacts is
given by the Landauer formula \mbox{$G=G_0\sum_{i=1}^{N}\tau_{i}
$}, where $\tau_i$ are the transmission probabilities for each of
$N$ conductance channels and $G_0=2e^2/h$ is the conductance
quantum. For a given contact realization the conductance channels
are in general neither completely open nor completely closed and
the transparency $\tau_i$ of each channel depends on the material
forming the contact \cite{Nature,Sirvent}, detailed atomic
arrangement and applied stress \cite{PRLplateau}. In this report
we study the channel transparency set $\{\tau_{i}\}$ both in gold
one-atom contacts and chains of single gold atoms \cite{Chains},
for which theoretical models
\cite{PRLplateau,Cuevas98,Lopez-Ciudad99,abinitio,Palacios02}
predict an almost completely open single channel, and therefore a
conductance close to $G_0$.

It is well established that conductance histograms (CH) show that
the conductance of one-atom contacts of gold is close to
$G_0$\cite{review} but the values of $\tau_i$ cannot be obtained
from the only measurement of the total conductance. However, the
marked non-linearity of the current-voltage characteristic
(\textit{IV}) of superconducting contacts has been exploited to
obtain the set of transmission coefficients $\{\tau_i\}$ of
atomic-sized aluminum contacts \cite{PRLset}. The channel
transmission probabilities are extracted by fitting the measured
\textit{IV} curve to a sum of $N$ independent \textit{IV}s
calculated for individual channels with a given transmission
probability \cite{Averin95,Cuevas96}. This method was afterwards
extended to other superconducting materials, showing that the
number of conducting channels contributing to the current in
one-atom contacts is limited by the valence of the atom at the
contact \cite{Nature}.

Gold is not superconducting. However we take advantage of the
superconducting correlations induced in a thin layer of normal
metal in contact with a superconductor. The energy spectrum is
modified and a gap is opened at the Fermi energy. At low voltages
transport is dominated by Andreev reflection processes which
results in non-linear \textit{IV} curves. Here we use the
aforementioned method \cite{PRLset,Nature} to analyze the
transmission coefficients of proximity induced superconducting
one-atom contacts and atomic chains of gold.

The proximity effect has been previously exploited
\cite{Nature,Scheerproxy} to get information of the $\{\tau_i\}$
in gold atomic contacts. In the first experiment \cite{Nature} the
\textit{IV} curves were fitted to a sum of theoretical
\textit{IV}s corresponding to BCS superconductors and the
contribution of a single channel in one-atom contacts is reported.
However, the energy dependence of the density of states and the
probability of Andreev reflection at a proximity induced
superconducting contact are smeared with respect to the BCS model
\cite{Belzig96}. The modifications due to the proximity character
of the superconducting correlations were taken into account in the
later experiment \cite{Scheerproxy}. Most of the experimental
\textit{IV}s recorded at the last conductance plateau before
contact breaking could be fitted with a single channel. Several
channels were necessary to fit the \textit{IV} curves of some
contacts with conductance smaller than $G_0$. However, in these
experiments the conductance of the smallest contacts is usually
much smaller than $G_0$, being the first peak of the histogram
placed at $0.6$ $G_0$ and plateaux not nearly flat. These results
are not consistent with what has been usually reported in gold
atomic contacts in the normal state \cite{review}. Here we report
experiments in which the conductance histograms for proximity
superconducting gold and normal gold are in remarkable agreement
(Fig 1). We also study the channel content of chains of gold
atoms.

Nanocontacts are produced by pressing two wires crosswise against
each other. The wires are used as electrodes in place of tip and
sample of an STM. The separation and contact size between the
wires can be controlled with the piezoelectric positioning system
the microscope. The advantage of using two crossed wires is the
possibility of selecting the position of the point contact along
both wires using the coarse lateral displacement capability of the
microscope, allowing for the exploration of point contacts at
different spots along the wires. The wires (0.25 mm diameter) are
made of bulk lead and are in a first preparation step covered by
thermal evaporation with a thick layer of lead (900 nm at a rate
of 0.8 nm/s). This thick layer of lead provides a clean surface. A
thin layer of gold (28 nm, rate of 0.1 nm/s) is then evaporated on
top of the Pb layer. The lead and gold deposition sequence is
performed without breaking the vacuum in the chamber ($<10^{-6}$
mbar), thus preventing the presence of oxide at the Pb-Au
interface and ensuring good electrical contact between both
layers. The substrate is at a temperature of 80 K during film
deposition. One-atom contacts are fabricated by slight indentation
($<3$ nm) and subsequent retraction of the electrodes. Further
retraction results in contact breaking and a jump to the
tunnelling regime. Experiments are done at 1.8 K. During
nanocontact pull-off the conductance trace is step-like (see Fig.
1). This characteristic behavior has been shown to be due to the
mechanical processes that take place during contact breaking
\cite{PRLforce}; conductance plateaus corresponding to elastic
deformation stages and sharp conductance changes related to sudden
rearrangements of the atoms in the narrowest part of the
nanocontact.

\begin{figure}
 \begin{center}
        \leavevmode
        \epsfxsize=85mm 
        \epsfbox{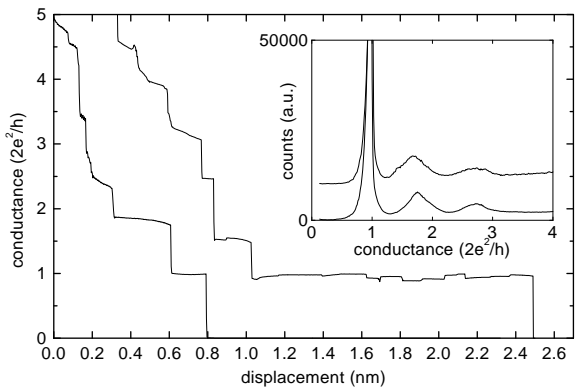}
 \end{center}
 \vspace{0mm}

 \caption{Differential conductance at a bias voltage of 10 mV of proximity induced superconducting gold contact during pull-off.
 One-atom contact formation (left) and atomic chain (right). The length of the conductance plateau indicates that
 a chain longer than five atoms is formed. Inset: conductance histograms of gold in the normal state (bottom)
 and proximity superconducting (shifted upwards 10000 counts).}
\label{FIG. 1.}
 \end{figure}

For gold nanocontacts in the normal state the last conductance
plateau before contact breaking is close to $G_0$ and corresponds
to the smallest possible contact: a one-atom contact. Despite the
inherent variability of the exact conductance trace during contact
pull-off, it has been shown in several experiments that the
conductance histogram (CH) over a large number of contact breaking
realizations displays peaks at conductance values that are
characteristic of the chemical nature of electrodes\cite{review}.
CH in gold nanocontacts have a characteristic first peak at a
conductance close to $G_0$. We show in the inset of Fig. 1 a
comparison between CH obtained for bulk gold tips in the normal
state and in proximity induced superconducting gold nanocontacts.
Due to the presence of excess current in the \textit{IV}s of
superconducting nanocontacts it is necessary to measure the
differential conductance at a fixed bias voltage well above
$\Delta/e$, where $\Delta$ is the energy of the superconducting
gap. The remarkable agreement between both CH supports the
validity of our analysis also for atomic-sized contacts of gold in
the normal state. The low value of the conductance of the first
peak in the conductance histogram in Ref. \onlinecite{Scheerproxy}
was probably due to enhanced elastic scattering related to sample
preparation method.

We show in Fig. 2 representative \textit{IV} curves recorded at a
one-atom contact, in an atomic chain and in the tunneling regime,
up to a voltage $2\Delta_0/e$, where $\Delta_0=1.40$ meV is the
bulk superconducting gap for lead. We show both the measured
\textit{IV} curves (symbols) and theoretical fitting (lines). The
theoretical \textit{IV}s in the contact regime are calculated by
solving the time-dependent Bogoliubov de Gennes equations within
the scattering formalism \cite{Averin95,Zaitsev97}. This method
requires the knowledge of the Andreev reflection amplitude of
probability ($a(E)$) at the contact, where the voltage drops.
$a(E)$ can be computed if the normal and anomalous Green functions
are known\cite{Zaitsev97}. In the tunneling regime the
\textit{IV}s are calculated by the usual convolution of the
densities of states at both sides of the barrier\cite{Tinkham}.

As both Au-Pb electrodes are only very weakly coupled at the weak
link, to calculate the density of states and the Andreev
reflection amplitude of probability, we model our system as two
independent normal-superconducting (NS) structures and solve
self-consistently the Usadel equations\cite{Usadel70,Belzig96}.
Usadel equations provide a quasiclassical description of the Green
functions of a superconductor in the dirty limit, in which
electronic transport is diffussive. Elastic impurity scattering is
included in the Born approximation and is characterized by a mean
free path $l$ (or a diffusion coefficient $D=v_Fl/3$). A
description based on Usadel equations was recently used to explain
the \textit{IV} curves of lead nanostructures under the influence
of a magnetic field and proximity effect, providing excellent
quantitative agreement with experiment both in the contact and
tunneling regime\cite{Suderow00,Suderow02}.

The two NS are assumed to be equal and consist of a dirty normal
layer with thickness $d_N$ which is bounded at one end by vacuum
and joined to a dirty semi-infinite superconductor at the other.
Note that the diffussive description is expected to be valid in
the case $d_N\ll l$. The Green functions which are relevant to
calculate the IV curves are the ones at the vacuum-bounded edge of
the normal layer. The proximity effect is also, affected by the
existence of a barrier at the interface, as quasiparticles
normally reflected do not contribute to it. In a diffusive system,
the mismatch between the characteristic parameters (conductivity
and diffusion coefficients) of the normal and superconducting
metals, leads to an effective barrier for the quasiparticles. It
can be described through the parameter
$\Gamma=(D_S/D_N)^{1/2}\sigma_N/\sigma_S$. In our model we assume
$\Gamma=1$ and a vanishing resistance of the
interface\cite{note1}. With these assumptions the superconducting
correlations in the normal metal are described by $\Delta_0$ and
the value of $d_N/\xi_S$, where $\xi_S$ is the coherence length of
the superconductor. In the atomic chain and one-atom contact the
transmission channels set enters also as fitting parameters. A
similar model was used by Scheer {\it et al}\cite{Scheerproxy} in
the  description of Au contacts with superconductiviy induced by
proximity effect. In their case, however, the mismatch parameter
$\Gamma$ was also used as a fitting parameter.

\begin{figure}
\vspace{0mm}
\begin{center}
        \leavevmode
        \epsfxsize=85mm
        \epsfbox{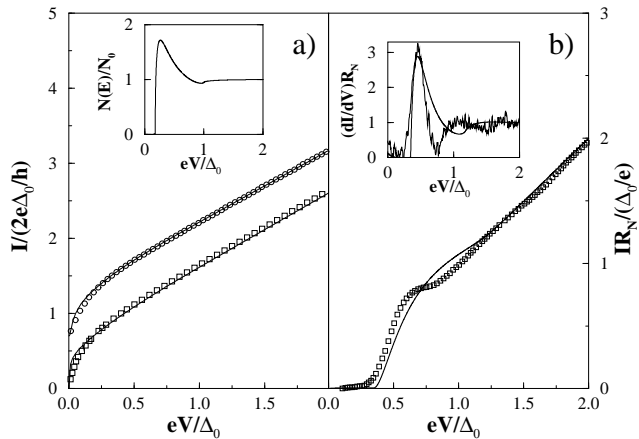}
 \end{center}
\vspace{0mm} \caption{Main figures show the \textit{IV} curves
corresponding to the tunneling regime (b), a single-atom contact
(bottom curve in (a)) and an atomic chain (top curve in (b)). The
curve corresponding to the atomic chain has been displaced for
clarity. Experimental results are shown by symbols and theoretical
fittings by solid lines. The transmissions obtained in the contact
regime are $T_1 = 0.995$ (single atom), and $T = 0.96$ (chain).
Inset in (a) shows the density of states used to calculate the
theoretical curves -see text. Inset in (b) show the derivative of
the experimental (the one with noise) and calculated tunneling
curves.} \label{FIG. 2.}
\end{figure}

Solid lines in Fig. 2 show the calculated \textit{IV} which for
the same sample fitting parameter, $d_N/\xi_S=0.81$, best fit both
the curves in the tunneling and the contact regime. Assuming the
nominal thickness $d_N=28$ nm it corresponds to a superconducting
coherence length $\xi_S=34$ nm, in good agreement with the values
obtained in Ref. \onlinecite{Suderow00} and
\onlinecite{Suderow02}.

The tunneling regime \textit{IV} curve, plotted in Fig. 2b, shows
a gap smaller than $\Delta_0$ and a bump characteristic of NS
structures \cite{Wolf82}. It is however poorly fitted
\cite{notefit}, which means that the density of states used, shown
in the inset of Fig. 2(a), does not agree completely with the
experimental one. The source of disagreement in the fitting can be
better understood by looking at the derivative of the experimental
and theoretical tunnel \textit{IV} curves, shown in inset in Fig.
2 (b). The theoretical curve shows an asymmetric peak at the gap
which decays slowly as the energy increases and results from the
one in the density of states. This asymmetric peak structure is
characteristic  of the diffusive regime for parameters which give
gaps in the normal metal much smaller than $\Delta_0$. The peak in
the experimental curve is more symmetric. More rounded peaks (and
tunneling curves with a bump more similar to the one obtained
experimentally) can be obtained for much smaller values of
$d_N/\xi_S$. In the diffusive regime the gap induced in the
spectrum of the normal metal decreases with $d_N/\xi_S$. Lower
values of the fitting parameter induce gaps with values much
closer to $\Delta_0$ and could not explain the strong reduction of
the gap found experimentally and would give worse fits.

Scheer {\it et al.} \cite{Scheerproxy} also found disagreement in
the fit of the tunneling conductance and related it to the
non-diffusive character of the samples, being the condition
$d_N\ll l$ not fulfilled. The proximity effect is a consequence of
the coherent superposition of Andreev reflection and is strongly
influenced by the length of the path which the electron travels
between Andreev reflection processes. Thus, the  energy dependence
of the induced pair correlations is very sensitive to the degree
of disorder in the normal metal and the shape of the spectrum
differs considerably in the clean and dirty
limits\cite{Arnold78,Wolf82,Belzig96}. We have also tried, without
success, to fit the experimental curves assuming that there is no
disorder being the transport in the electrodes ballistic, instead
of diffusive\cite{Arnold78,Wolf82,Kieselmann87,Vecino}. As
concluded by  Scheer {\it et al.} \cite{Scheerproxy}, we think
that the lack of good fittings in the tunneling regime  is due to
transport in the gold layer being in the weakly disordered regime,
instead of in the diffusive or clean limits.

Given the uncertainty in the density of states, the accuracy with
which the channel transmission set is obtained in the contact
regime is slightly reduced, compared to the ones done in BCS
superconductors. The single-atom curve can be reasonably well
fitted with a very open channel with transmission $T_1=0.995$. The
chain is reasonably well fitted by a single channel with
transmission $T = 0.96$. This result is representative of the
general behavior that we observe thus confirming that one widely
open channel is responsible for the conduction in single atom gold
contacts, in agreement with theoretical
predictions\cite{Cuevas98,Lopez-Ciudad99,abinitio,Palacios02}.

This work has been supported by the Spanish DGI through grants
MAT2001-1281 and PB0875/96. Additional funding comes from the ESF
Vortex programm, the Welch Foundation, and the NSF under grants
DMR-0210383 and DMR0115947. E.B. acknowledges financial support
from a fellowship of the Comunidad Aut\'onoma de Madrid during the
early stages of this work. We acknowledge W. Belzig for helpful
support.

$^*$ Present  address: Theoretische Physik, ETH-H\"onggerberg,
CH-8093 Z\"urich, Switzerland.


\begin{thebibliography}{99}
\bibitem{note1}
We have found that to introduce the influence of the mismatch of
the Fermi velocities and mean free path in gold and lead or of an
interface resistance does not improve significantly the quality of
the fittings and would introduce more fitting parameters.
\end{thebibliography}
\end{document}